# Operational risk modeled analytically II: the consequences of classification invariance

Vivien BRUNEL

Head of Risk and Capital Modeling, Société Générale
and
Professor of Finance, Léonard de Vinci Pôle Universitaire, Finance Lab, Paris La Défense, France

vivien.brunel@socgen.com

(this version: May 11th, 2015)

Most of the banks' operational risk internal models are based on loss pooling in risk and business line categories. The parameters and outputs of operational risk models are sensitive to the pooling of the data and the choice of the risk classification. In a simple model, we establish the link between the number of risk cells and the model parameters by requiring invariance of the bank's loss distribution upon a change in classification. We provide details on the impact of this requirement on the domain of attraction of the loss distribution, on diversification effects and on cell risk correlations.

**Introduction**

It is critical to have a robust and sound methodology for operational risk assessment and capital measurement. The most common approach (namely the Loss Distribution Approach) for measuring operational risk within a bank is based on the frequency and severity estimation from observed events. However, it is challenging to build accurate and robust estimations in this framework, as shown by Cope *et al.* (2009).

One major issue of the LDA approach is to pool observed events based on a risk and business line classification. It is well established that banks having similar risk profiles will select different classifications. The choice of the classification is critical regarding the parameters and the outputs of the model: a change in the classification may result in a significant change in the outputs of the model, even though the total risk of the bank should not depend on the very details of the classification but rather on its global risk profile. For instance, when a bank performs an internal re-organization of its analytical structure without changing anything to its global risk profile, the resulting change of the cell definitions in the model may lead to unexplained impacts.

This drawback is widely accepted for lack of anything better but it is not sound from the regulator or from the bank's management perspective that the model provides significant differences in the outputs when we change the classification. Restoring this coherence has never been studied in the literature as far as we know, and it is a complex issue. Firstly, it is impossible to require strict classification invariance for the bank's loss distribution because nobody knows the real bank's loss distribution. Secondly, some of the statistical indicators for the bank's loss, such as the expected loss or the variance of the loss, are known with convenient reliability; asking for classification invariance for these indicators only, would generate additional constraints upon the cell parameters which would require to be estimated jointly and no longer separately at cell level.

When designing its operational risk model, the bank justifies the choice of classification based on expert and statistical evidences. The choice of the risk categories depends on the processes in place within the bank to monitor or manage operational risk. Risk categories definitions or perimeter may change over time, in particular when emerging risks arise, or when the exposure to certain types of risk changes dramatically. The business line definitions are subject to some changes as well, as soon as the bank operates an internal reorganization, launches, buys or sells a line of business. Another critical choice for the bank is to determine the number of cells to be included in the model. Modeling is only possible when cells represent somewhat "homogeneous risks", but the granularity of the modeling is constrained by the amount of data available on each cell for calibration. To the best of our knowledge, the impact of the number of cells on the bank's loss distribution, on the diversification benefits or on the correlations between cells is still an open question and has never been studied so far.

In this paper, we study the impact of the classification invariance requirement on a class of operational risk models described directly by the total loss distribution at cell level instead of frequencies and individual loss severities as it is usually the case in the LDA framework. Analytical results emerge in the limit of an infinite number of cells. We show, in the case of homogeneous cell risks, that the parameters of the cell loss distributions scale with the number of cells, which is the first time that a link between the model structure and the parameters is established. To get this link, we apply some recent results of the mathematical theory (Ben Arous *et al.*, 2005; Bovier *et al.*, 2002) to the asymptotic regime in operational risk models. Our main results are:

- The bank's loss distribution belongs to the domain of attraction of the fully asymmetric Lévy distribution in all realistic cases (having a variance of the total bank's loss equal to zero is not realistic).
- Increasing the number of cells does not necessarily generate a lower capital charge or a higher diversification benefit.
- Increasing the number of cells generates negative diversification and, at the same time, decorrelation among cell risks.

The structure of the paper is as follows. In section 1 we study the specific case of lognormal cell losses. We show in particular that the bank's loss distribution is not in the domain of attraction of the normal distribution. In section 2, we study the impact of classification invariance in the asymptotic limit in the general case. We focus on diversification effects in section 3 and on correlations between cell risks in section 4.

### 1. Lognormal case

We consider the bank's overall operational risk as a portfolio of $N$ operational risks at cell level. We suppose in this paper that these risks are all independent and identical to each other (i.e. have the same repartition function). In this section, we assume that cell losses $(Y_i)_{1 \leq i \leq N}$ are independent and identically distributed (*i.i.d.*) lognormal random variables. The bank's loss is the sum of the individual cell losses:

$$L_N = \sum_{i=1}^{N} Y_i = \sum_{i=1}^{N} e^{\mu_N + \sigma_N X_i} \tag{1}$$

where the risk parameters $\mu_N$ and $\sigma_N$ depend on the number of cells, and $(X_i)_{1 \leq i \leq N}$ are i.i.d. normal random variables. Classification invariance implies that the expected value and variance of the total loss converge to a constant value when the number of cells $N$ goes to infinity. This leads to:

$$\begin{aligned}\lim_{N \to \infty} E[L_N] &= \lim_{N \to \infty} N\, e^{\mu_N + \sigma_N^2/2} = a \\ \lim_{N \to \infty} var[L_N] &= \lim_{N \to \infty} N\, e^{2\mu_N + \sigma_N^2}\left(e^{\sigma_N^2} - 1\right) = b\end{aligned} \tag{2}$$

In the limit $N \to \infty$, eq. (2) leads to the following scaling relation of the risk parameters with the number of cells:

$$\mu_N \sim -\frac{3}{2} \ln N \quad \text{and} \quad \sigma_N \sim \sqrt{\ln N} \tag{3}$$

The Lindeberg condition defines the domain of attraction of the normal distribution (see Feller, 1957):

$$\lim_{N\to\infty} \frac{1}{s_N^2} \sum_{k=1}^{N} \int_{|Y_k - E(Y_k)| > \varepsilon s_N} [Y_k - E(Y_k)]^2 d\mathbb{P} = 0 \text{ for all } \varepsilon > 0 \quad (4)$$

Bovier *et al.* (2002) show from eq. (4) that the domain of attraction of the normal distribution for the sum of lognormal random variables is defined by $\sigma_N < \sqrt{\frac{1}{2}\ln N}$. We conclude that the limiting law of $L_N$ - defined by eq. (1) and (3) - is not Gaussian as $\sigma_N$ scales faster than $\sqrt{\frac{1}{2}\ln N}$, in spite of finite first and second moments.

## 2. General case

### 2.1. Limit theorem for sums of random exponentials

Section 1 illustrates that requiring classification invariance leads to unexpected conclusions in the lognormal case. A general theorem by Ben Arous *et al.* (2005) (BBM theorem hereafter) describes the domains of attraction for sums of random exponentials in the general case. In this sub-section, we study the limit distribution of the sum $S_N(t) = \sum_{i=1}^{N} e^{tX_i}$ where $(X_i)_{1 \le i \le N}$ is a sequence of *i.i.d* random variables whose repartition function is regularly varying at infinity with index $\rho > 1$. The repartition function $H_\rho$ of each random variable $X_i$ behaves as the Weibull distribution at infinity, *i.e.* $H_\rho(x) \sim 1 - \exp(-cx^\rho)$ for $x \to \infty$. Of course, the Weibull distributions belong to this class of distributions and the normal distribution as well, with an index equal to $\rho = 2$.

When the parameter $t$ is a constant, the usual Central Limit Theorem applies for all $\rho > 1$. Indeed, the variance of $e^{tX_i}$ is finite, and the sum $S_N(t)$ belongs to the domain of attraction of the normal distribution. On the other hand, when the parameter $t$ goes to infinity, the Central Limit Theorem no longer applies and the sum $S_N(t)$ may belong to other domains of attraction as illustrated in section 1. Ben Arous *et al.* have exhibited the limiting behavior of the sum $S_N(t)$ when both $N$ and $t$ go to infinity.

Let's introduce the following notations:

- the modified index value $\rho' = \rho/(\rho - 1)$
- the parameter $\lambda$ measuring the speed at which $t$ goes to infinity with $N$ such that $t \sim \left(\frac{\rho'}{\lambda}\ln N\right)^{1/\rho'}$
- $\alpha = (\rho\lambda/\rho')^{1/\rho'}$

By applying BBM theorem (cf. theorems 2.1 - 2.3 in Ben Arous *et al.*, 2005), we get:

$$\frac{S_N(t) - A(t)}{B(t)} \xrightarrow{d} \begin{cases} \mathcal{F}_\alpha & \text{if } \alpha < 2 \\ N(0,1) & \text{if } \alpha \ge 2 \end{cases} \quad (5)$$

where $\mathcal{F}_\alpha$ is the fully asymmetric ($\beta = 0$) Lévy distribution, $N(0,1)$ is the normal distribution with mean 0 and variance 1, and:

$$A(t) = \begin{cases} E[S_N(t)] & \text{if } \alpha > 1 \\ \frac{1}{2}E[S_N(t)] & \text{if } \alpha = 1 \\ 0 & \text{if } \alpha < 1 \end{cases} \text{ and } B(t) = \begin{cases} var[S_N(t)] & \text{if } \alpha > 2 \\ \frac{1}{2}var[S_N(t)] & \text{if } \alpha = 2 \\ N^{\rho/\alpha} & \text{if } \alpha < 2 \end{cases} \quad (6)$$

If we call $H(t) = \ln E[e^{tX}]$, Ben Arous *et al.* (2005) show that $H(t) \sim t^{\rho'}/\rho'$ (we have the strict equality in the case of the Weibull distribution). Then we get:

$$E[S_N(t)] = Ne^{H(t)} \sim N^{\frac{\lambda+1}{\lambda}} \qquad (7)$$

In the lognormal case (i.e. when the $(X_i)_{1 \leq i \leq N}$ are normal random variables), we have $\rho = \rho' = 2$. The classification invariance condition of eq. (3) leads to $\lambda = 2$ and $\alpha = \sqrt{2}$, and then we get $E[S_N(t)] \sim N^{3/2}$.

2.2. Asymptotic classification invariance in operational risk models

If we assume that the bank's operational loss is the sum of random exponentials, we can write it, as explained in section 1:

$$L_N(t) = e^{\mu_N} S_N(t) \qquad (8)$$

Where $e^{\mu_N}$ is a parameter depending on the number of cells only and $t = \sigma_N$. BBM theorem still applies for the bank's loss $L_N(t)$ just by replacing $S_N(t)$, $A(t)$ and $B(t)$ by $L_N(t)$, $e^{\mu_N}A(t)$ and $e^{\mu_N}B(t)$ respectively in eq.(5). Depending on the value of $\alpha$, BBM theorem states that we have three different regions in the parameters space, which are separated to each other by two curves plotted in figure 1 and corresponding to $\alpha = 1$ and $\alpha = 2$ respectively. When expressed in terms of the scaling parameter $\lambda$ and of the cell loss tail parameter $\rho$, those curves are defined by the following equations:

$$\begin{aligned} \alpha = 1 &\iff \lambda = \frac{1}{\rho-1} \qquad \text{(Curve A)} \\ \alpha = 2 &\iff \lambda = \frac{2\rho'}{\rho-1} \qquad \text{(Curve B)} \end{aligned} \qquad (9)$$

When $\alpha > 2$ (above curve B), the Central Limit Theorem (CLT) still applies. Conversely, when $1 < \alpha < 2$ (under curve B), the Central Limit Theorem no longer holds but the Law of Large Numbers (LLN) remains because $\frac{L_N(t)}{E[L_N(t)]} \xrightarrow{p} 1$ when the number of cells goes to infinity as shown in theorem 2.1 of Ben Arous *et al.* (2005). We notice from eq. (6), for any value $\alpha < 2$, the unusual scaling behavior $B(t) = N^{\rho/\alpha}$ compared to $N^{1/\alpha}$ when the Central Limit Theorem applies for sums of *i.i.d.* fat tailed random variables. Finally, when $\alpha < 1$ (under curve A), both the Central Limit Theorem and the Law of Large Numbers break down.

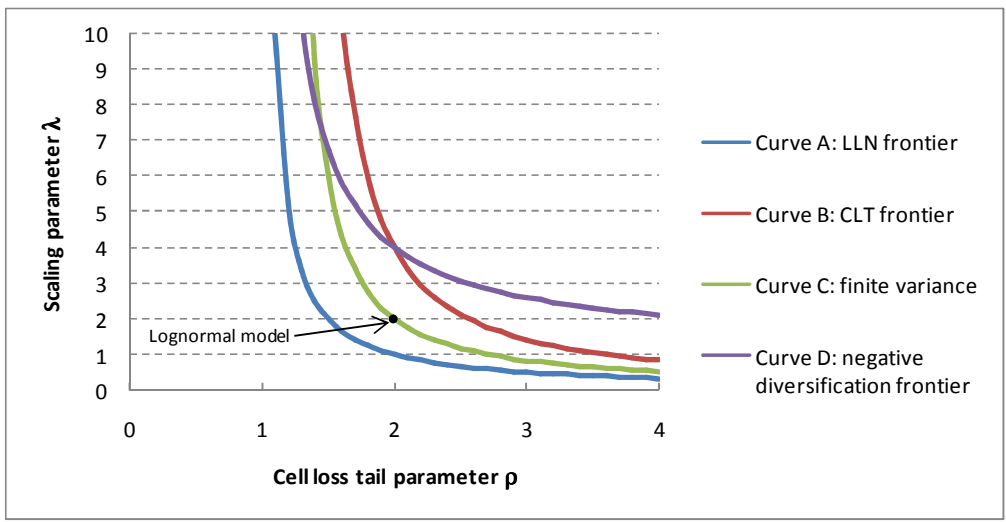

*Figure 1: Limiting behavior of operational risk models as a function of the scaling parameter λ and of the cell loss tail parameter ρ.*

Asymptotic classification invariance leads to the requirement that the expected loss is finite and is independent from the number of cells:

$$\lim_{N \to \infty} e^{\mu_N} A(t) = O(1)$$

From eq. (7), this condition leads to the classification invariance condition:

$$\mu_N \sim -\frac{\lambda+1}{\lambda} \ln N \tag{11}$$

Additionally, as we have $var(e^{tX}) \sim E[e^{2tX}] \sim e^{H(2t)}$, the variance of the bank's loss writes:

$$var(L_N(t)) = N e^{2\mu_N} var(e^{tX_i}) \sim N^{1-2\frac{\lambda+1}{\lambda}+\frac{2\rho'}{\lambda}} \tag{12}$$

A finite variance for the bank's loss is only achieved when the exponent is equal to 0, i.e.:

$$\lambda = 2\rho' - 2 \qquad \text{(Curve C)} \tag{13}$$

In all the other cases, the variance is either infinite or equal to 0, depending on the sign of the exponent in eq. (12). In figure 1, curve C corresponds to models with a finite variance as stated in eq. (12). The region corresponding to an infinite variance (resp. variance equal to 0) is located under (resp. above) curve C. We observe that, for any value of the parameter $\rho$, finite variance models correspond to a value of $\alpha$ in the range between 1 and 2: surprisingly, all realistic models under the classification invariance requirement (either with finite or infinite variance) belong to the domain of attraction of the fully asymmetric Lévy distribution and not of the normal distribution. Additionally, we notice that finite variance models correspond to one curve in the parameters space, and are an infinitely small proportion of all the possible models.

Back to the lognormal case under classification invariance with finite variance, eq. (6) and (11) lead to $e^{\mu_N} B(t) \sim N^{\sqrt{2}-3/2}$. If we set $\epsilon_N = N^{3/2-\sqrt{2}}(L_N(t) - E[L_N(t)])$, BBM theorem states that $\epsilon_N \xrightarrow{d} \mathcal{F}_{\sqrt{2}}$. As the bank's loss $L_N(t)$ has a finite variance, we obtain that the variance of $\epsilon_N$ diverges as $N^{3-2\sqrt{2}}$ when $N \to \infty$. For any arbitrary value of the parameter $\rho > 1$, eq. (6) and (11) show that the variance of $\epsilon_N$ diverges as $N^{2\left(\frac{\lambda+1}{\lambda}-\frac{\rho}{\alpha}\right)}$.

### 3. Diversification impact

The requirement of classification invariance generates non trivial impacts on diversification benefits. We measure these impacts with the diversification ratio, which is equal to the ratio between the quantile of the bank's loss and the sum of the loss quantiles at cell level:

$$DR = \frac{VaR_q(L_N(t))}{\sum_{i=1}^{N} VaR_q(Y_i)} \tag{14}$$

Neslehova *et al.* (2006) showed that fat-tailed (tail index parameter lower than 1) individual cell losses $Y_i$ would result in negative diversification ($DR > 1$). Brunel (2014) showed that sub-additivity of VaR could disappear in case of the sum of non fat-tailed random variables, for instance when these random variables are correlated lognormal variables. From eq. (8), we get the sum of stand-alone values at risk:

$$\sum_{i=1}^{N} VaR_q(Y_i) = N e^{\mu_N + t H_\rho^{-1}(q)}$$

From eq. (5), we get the bank's capital charge:

$$VaR_q(L_N(t)) = E[L_N(t)] + e^{\mu_N} B(t) \mathcal{F}_\alpha^{-1}(q)$$

where the function $\mathcal{F}_\alpha(\cdot)$ is the repartition function of the fully asymmetric Lévy distribution $\mathcal{F}_\alpha$ with tail exponent $\alpha$. The diversification ratio is then equal to:

$$DR \sim \mathcal{F}_\alpha^{-1}(q).\exp\left[\left(\frac{\rho}{\alpha} - 1\right)\ln N + H_\rho^{-1}(q)\left(\frac{\rho'}{\lambda}\ln N\right)^{1/\rho'}\right] \quad (15)$$

As $1/\rho' < 1$, the dominant term in the exponential is always the first one for $N$ sufficiently large. The behavior of the diversification benefit depends on the tail index $\alpha$ of the bank's aggregate loss compared to the tail index $\rho$ of the individual losses. When $\alpha > \rho$, the diversification ratio decreases with $N$, and we have a positive diversification of risk. On the reverse, when $\alpha \leq \rho$, the diversification index increases with $N$ and we have negative diversification. We plot the frontier between these two regimes in figure 1 (curve D), whose equation writes:

$$\alpha = \rho \iff \lambda = \frac{\rho \rho'}{\rho - 1} \quad \text{(Curve D)} \quad (16)$$

In the lognormal model, as $\rho = 2$ and $\alpha = \sqrt{2}$, we see that we have negative diversification between cells in the asymptotic limit for $N$ large enough. Increasing the number of cells in the model does not necessarily result in a lower capital charge or a higher diversification benefit. This result is new compared to those of Neslehova *et al.* (2006) or Brunel (2014), we show here that super-additivity may occur in operational risk models when both the first and second moments of the loss distributions at cell level and at the bank level are finite. Moreover, it is surprising to observe in figure 1 that negative diversification may even occur in some areas of the parameter space that belong to the domain of attraction of the normal distribution ($\alpha > 2$).

4. **Correlations between cells**

Let's come back to the lognormal case with a bank's loss having a finite variance and consider the one factor model with uniform correlations. The bank's loss writes in this case (see Brunel, 2014):

$$L_N = \sum_{i=1}^{N} e^{\mu_N + \sigma_N(\sqrt{\rho_N}F + \sqrt{1-\rho_N}X_i)} = e^{\mu_N + \sigma_N\sqrt{\rho_N}F} \sum_{i=1}^{N} e^{\sigma_N\sqrt{1-\rho_N}X_i}$$

where $F$ is a common normal risk factor for all cells, independent of the $(X_i)_{1 \leq i \leq N}$, and $\rho_N$ is the correlation parameter between cells. As this correlation parameter is not larger than 1, the classification invariance conditions don't change, and the parameter $\sigma_N$ still scales as $\sqrt{\ln N}$ and $\mu_N$ scales as $-3/2 \ln N$. As a consequence, the bank's loss remains equal to $O(1)$ only if $\sigma_N\sqrt{\rho_N} = O(1)$. We get the following scaling behavior of the correlation parameter:

$$\rho_N \propto \frac{1}{\ln N} \quad (17)$$

This suggests that classification invariance may lead to risk independence in the asymptotic limit. This is the reason why the independence assumption of sections 1 to 3 deems relevant to study operational risk portfolios.

**Discussion and conclusion**

As far as we know, this paper is the first attempt of exploring operational risk models under the requirement of classification invariance. The results we obtain are unexpected and may challenge some of the knowledge we have regarding operational risk models. Indeed, we show that operational risk models with a finite variance of the bank's loss all belong to the domain of attraction of the fully asymmetric Lévy distribution; models that belong to the normal distribution domain of attraction all have a loss variance equal to 0 in the asymptotic limit. Moreover negative diversification may occur when the number of cells is large enough; this is in particular the case with the lognormal model, which is a surprise. Finally, when cell risks are correlated, the above results are unchanged and classification invariance generates decorrelation between cells when the number of cells increases.

Further research is necessary to explore in more details the negative diversification effects, to assess for instance if we enter the "one loss causes ruin" regime of Neslehova *et al* (2006)*.* The classification invariance requirement for real operational risk portfolios is an unexplored field, in particular regarding parameters estimation for models having different risk profiles at cell level.